\documentclass[runningheads]{llncs}

\usepackage[T1]{fontenc}
\usepackage{lmodern}
\usepackage{graphicx}
\usepackage{amsmath,amssymb}
\usepackage{booktabs}
\usepackage{array}
\usepackage{multirow}
\usepackage{placeins}
\usepackage{tabularx}
\usepackage{url}

\newcommand{\CER}{\operatorname{CER}}
\newcommand{\Edit}{\mathcal{E}_{\mathrm{char}}}
\newcommand{\ReadScore}{\operatorname{ReadScore}}
\newcommand{\dur}{\operatorname{dur}}
\newcommand{\ind}{\mathbb{I}}

\begin{document}

\title{Capability-Guided Iterative Online Optimization of Cascaded
Spoken-Dialogue Modules}
\titlerunning{Capability-Guided Online Optimization}

\author{Yuhang Qiu \and Bohan Li \and Wenbin Huang \and Zhihan Li \and
Yiwei Guo \and Hankun Wang \and Xie Chen \and Kai Yu\inst{*}}
\authorrunning{Y. Qiu et al.}
\institute{X-LANCE Lab, Department of Computer Science and Engineering, \mbox{Shanghai Jiao Tong University}\\
MoE Key Lab of Artificial Intelligence, \mbox{Jiangsu Key Lab of Language Computing, China}\\
\textsuperscript{*}Corresponding author: Kai Yu}

\maketitle

\begin{abstract}
Cascaded spoken dialogue systems comprise perception-and-comprehension,
decision-making, and response-generation modules. We study iterative
optimization of the first and last modules. Their reciprocal speech--content interfaces enable one-sided data
to provide cross-module supervision: acoustic observations, such as waveforms
or feature sequences, can be pseudo-labeled to train generation, whereas
content-only samples can be rendered into speech to train comprehension.
Because the reliable teacher--student direction varies across domains, we
propose a two-stage, capability-guided pipeline for iterative online
optimization.
Stage~1 compares both modules through a shared intelligibility probe and
obtains a domain-wise capability prior. Stage~2 activates the compatible
one-sided-data route, constructs and filters supervision with the current
teacher, updates the selected student, and refreshes the relevant checkpoints
across cycles. We instantiate comprehension with Whisper large-v3 and
generation with CosyVoice2. After only 200 steps for Whisper and 600 for CosyVoice2 of our iterative online optimization, we observe that, on the Chinese MAGICDATA-RAMC dataset, Whisper reduces CER from 14.18\% to 10.77\%, while CosyVoice2 reduces it from 4.78\% to 4.04\%. On the Japanese Common Voice dataset, Whisper reduces CER from 15.23\% to 15.10\%, while CosyVoice2 reduces it from 14.85\% to 11.18\%. These results demonstrate the feasibility of
capability-guided iterative optimization of complementary spoken-dialogue
modules.

\keywords{Cascaded spoken dialogue systems \and Capability-guided optimization
\and Online optimization \and Speech recognition and synthesis}
\end{abstract}

\section{Introduction}

Cascaded spoken dialogue systems conventionally decompose speech interaction
into three functional modules. A perception-and-comprehension module maps an
acoustic observation, such as a speech waveform or acoustic-feature sequence,
to a content representation; a decision-making module interprets that content
and determines a response; and a response-generation module realizes the
response as an acoustic signal. ASR, a language model, and TTS are common
implementations of these three roles.

Modular spoken-dialogue frameworks such as X-Talk connect speech front ends
and understanding models with LLM-based decision-making and speech generation
in a cascaded speech-to-speech interface \cite{liu2025xtalk}. This modularity
allows the peripheral comprehension and generation modules to be evaluated and
optimized independently while leaving the central decision-making policy
unchanged. We focus on the iterative improvement of these two peripheral
modules.

The two modules provide approximately reciprocal, rather than invertible,
speech--content interfaces. A comprehension module $C$ maps an acoustic
observation $y$ to content $x$, whereas a generation module $G$ renders content
$x$ and acoustic context $p$ as speech. Transcription discards speaker,
prosodic, and environmental variation, while the same content admits multiple
valid speech realizations. Consequently, speech-only data can be pseudo-labeled
by $C$ and used with the original speech targets to train $G$. Conversely,
content-only data can improve $C$ after being rendered into speech by $G$.
Speech-chain methods exploit this duality to learn ASR and TTS from unpaired
data \cite{tjandra2017speechchain,ren2019almost}, while TTS-to-ASR pipelines
construct synthetic speech for semi-supervised ASR
\cite{baskar2019semisupervised,qi2023speechain}. Their teacher--student update
schedules are generally fixed in advance.

A predefined schedule for teacher--student updates can be suboptimal when
module strengths vary across language, domain, and acoustic condition. A comprehension model
may provide more trustworthy labels in one domain, whereas a generation model
may produce more intelligible speech in another. Using the weaker module
indiscriminately as teacher transfers its errors to the counterpart.
Materializing synthetic supervision once also fixes both the teacher and the
selected examples even though the modules change during optimization. These
issues motivate capability-guided routing followed by iterative online
supervision construction.

Our pipeline has two stages. During Stage~1, \emph{advantage-prior
acquisition}, both modules are evaluated on held-out domain probes through a
shared intelligibility measure. The resulting domain-wise prior, together with
the available one-sided data, determines the module to update. During Stage~2,
\emph{iterative online optimization}, the current teacher constructs
supervision, route-specific rules filter the candidates, and the accepted data
update the selected student. When $C$ is advantaged, speech-only samples are
pseudo-labeled by $C$ and used to update $G$. When $G$ is advantaged,
content-only samples are rendered into speech by $G$, filtered, and used to
update $C$.

Optimizer and learning-rate states continue across cycles, while later
supervision is constructed with newly resolved checkpoints. The supervision
distribution can therefore differ from that of a corpus constructed once
before training. Bilateral OPD constraints against the original modules limit
deviation during the selected updates. All capability, routing, selection,
checkpoint-mapping, and stopping rules are fixed rather than learned.

We instantiate $C$ with Whisper large-v3 and $G$ with CosyVoice2. In the
comprehension-supervised route, ASR pseudo-transcribes real Japanese speech; a
round-trip intelligibility check selects pairs for TTS training, whose acoustic
target remains the original waveform. In the generation-supervised route, TTS
synthesizes Mandarin speech from source text; duration, student-side
intelligibility, and ReadScore checks select pairs for ASR training. The fixed
pipeline executes both routes and returns the optimized ASR/TTS pair.

Our contributions are threefold:

\begin{itemize}
  \item We formulate a two-stage capability-guided pipeline that obtains a
  domain-wise module-advantage prior and selects whether to update
  comprehension or generation from the compatible one-sided data.
  \item We develop two reciprocal iterative online optimization modes with
  route-specific supervision construction and filtering, checkpoint refresh
  across cycles, continuous optimizer and learning-rate states, and bilateral
  OPD preservation.
  \item We instantiate the pipeline with Whisper large-v3 and CosyVoice2.
  Relative to its official initialization, each optimized module reduces CER
  on both the Japanese and Mandarin evaluations. Under matched step budgets
  and learning-rate schedules, iterative online Whisper outperforms its
  one-shot static reference on both tests, whereas iterative online CosyVoice2
  is slightly worse than its static reference on Japanese and trails it on
  Mandarin.
\end{itemize}

\section{Related Work}

\subsection{Joint ASR--TTS Optimization from Unpaired Data}

Machine speech-chain methods exploit the duality between speech recognition
and synthesis. Tjandra et al. connect both directions: on speech-only data,
ASR provides text for TTS reconstruction, while on text-only data, TTS
provides speech for ASR reconstruction \cite{tjandra2017speechchain}. Ren et al. combine denoising
auto-encoding, dual transformation, and bidirectional sequence modeling to
improve both tasks with a small paired set and additional unpaired speech and
text \cite{ren2019almost}. These approaches establish reciprocal supervision,
but their learning directions and coupled objectives are defined in advance.
Our pipeline instead uses domain-wise capability assessment to select the
teacher--student direction before iterative online optimization.

\subsection{Synthetic-Speech Supervision for ASR}

Baskar et al. combine ASR-to-TTS and TTS-to-ASR reconstruction objectives to
use unpaired speech and text for semi-supervised sequence-to-sequence ASR
\cite{baskar2019semisupervised}. SpeeChain scales the TTS-to-ASR direction with
multi-GPU synthesis, mixed real and synthetic batches, and on-the-fly data
selection \cite{qi2023speechain}. Both demonstrate the value of generated
speech for recognition, but focus on a prescribed ASR update route. Our
pipeline supports both reciprocal directions and uses a domain-level
capability prior to select the updated module.

\subsection{Modular Speech and Spoken-Dialogue Pipelines}

Modular speech systems increasingly compose specialized models as callable
components. AudioGPT uses ChatGPT to coordinate foundation models for speech,
music, and sound understanding and generation, with ASR and TTS providing the
spoken interaction interface \cite{huang2024audiogpt}. Speech-Copilot builds
modular speech-processing toolsets through task decomposition and uses an LLM
to generate programs that compose these tools
\cite{kuan2024speechcopilot}. At the spoken-dialogue level, AURA combines
open-weight ASR, an LLM, and TTS in a cascade in which the LLM dynamically
invokes external services during multi-turn voice tasks
\cite{maben2025aura}. These systems organize deployed speech components at
task time. Our focus is complementary: we study iterative training of the
comprehension and generation modules within a modular cascade while leaving
the decision-making and interaction policy unchanged.

\section{Method}
\label{sec:method}

\subsection{Capability-Guided Iterative Optimization Pipeline}

Consider a cascaded spoken dialogue system with an initial
perception-and-comprehension module $C_0$, an initial response-generation
module $G_0$, and a decision-making module between their interfaces. The
decision-making module and dialogue policy remain unchanged. The comprehension
interface maps an acoustic observation $y$, such as a waveform or feature
sequence, to content $x$, whereas the generation interface renders content
$x$ and acoustic context $p$ as speech. Candidate domains are denoted by
$\mathcal D=\{D_1,\ldots,D_M\}$, where $M$ is the number of domains. The
pipeline applies when the two interfaces are modality-dual and can be compared
through a common task-level intelligibility probe.

The pipeline operates on the one-sided pools
\begin{equation}
\mathcal U_D^{\mathrm{speech}}=\{y_i\},\qquad
\mathcal U_D^{\mathrm{content}}=\{(x_j,p_j)\}.
\end{equation}
Here $\mathcal U_D^{\mathrm{speech}}$ is the speech-only pool, and
$\mathcal U_D^{\mathrm{content}}$ is the content-only pool with generation
context. Stage~1 computes a domain-wise capability prior and selects an update
direction. Stage~2 repeatedly constructs supervision, applies route-specific
selection, updates the selected module, and resolves the relevant checkpoints.
A comprehension advantage activates $\mathcal U_D^{\mathrm{speech}}$ to
update $G$. A generation advantage causes $G$ to render
$\mathcal U_D^{\mathrm{content}}$ into speech, and the accepted synthetic
speech--content pairs update $C$. The pipeline state comprises the capability
prior, active routes, current module checkpoints, optimizer and scheduler
states, and stopping budgets. The final optimized pair
$(C^{\star},G^{\star})$ is returned when those budgets are reached.

\subsection{Capability-Advantage Assessment and Routing}

\subsubsection{Intelligibility-based probe and routing rules.}
Stage~1 evaluates the two modules and applies the capability-based routing
rule. From this point onward, we implement $C$ with ASR and $G$ with TTS. For
a speech domain $D$, let the held-out probe be
\begin{equation}
\mathcal P_D=\{(x_i,y_i,p_i)\}_{i=1}^{N_D},
\end{equation}
where $x_i$ is reference text, $y_i$ is a real waveform, $p_i$ is the prompt
waveform supplied to TTS, and $N_D$ is the number of probe utterances. The
prompt $p_i$ is acoustic context: depending on the corpus, it is either a
same-utterance segment extracted by voice activity detection (VAD) or a
preselected reference from a speaker pool.

For reference strings $\mathcal R=\{r_i\}$ and hypotheses
$\mathcal H=\{h_i\}$, corpus CER is
\begin{equation}
\CER(\mathcal R,\mathcal H)
=\frac{\sum_i(S_i+D_i+I_i)}{\sum_i|r_i|},
\label{eq:cer}
\end{equation}
where $S_i$, $D_i$, and $I_i$ are the character-level substitution, deletion,
and insertion counts after corpus-specific text normalization, and $|r_i|$ is
the normalized reference length.

The comprehension-side error is
\begin{equation}
E_C(D)=\CER\bigl(\{x_i\},\{C_0(y_i)\}\bigr),
\label{eq:comprehension-capability}
\end{equation}
where $C_0(y_i)$ is the transcript predicted from waveform $y_i$ and $E_C(D)$
is the comprehension error on $D$. To measure generation intelligibility, we
synthesize and then transcribe:
\begin{equation}
\tilde y_i=G_0(x_i,p_i),\qquad
\tilde x_i=J_D(\tilde y_i),
\label{eq:generation-eval-chain}
\end{equation}
\begin{equation}
E_G(D)=\CER\bigl(\{x_i\},\{\tilde x_i\}\bigr).
\label{eq:generation-capability}
\end{equation}
Here $\tilde y_i$ is synthetic speech, $J_D$ is the frozen ASR judge for
domain $D$, $\tilde x_i$ is the judge transcript, and $E_G(D)$ is the
generation-side intelligibility error. The two errors share the same probe and
normalization, while $J_D$ is selected independently of the comprehension
module $C_0$.

\begin{figure}[!t]
  \centering
  \includegraphics[width=\linewidth]{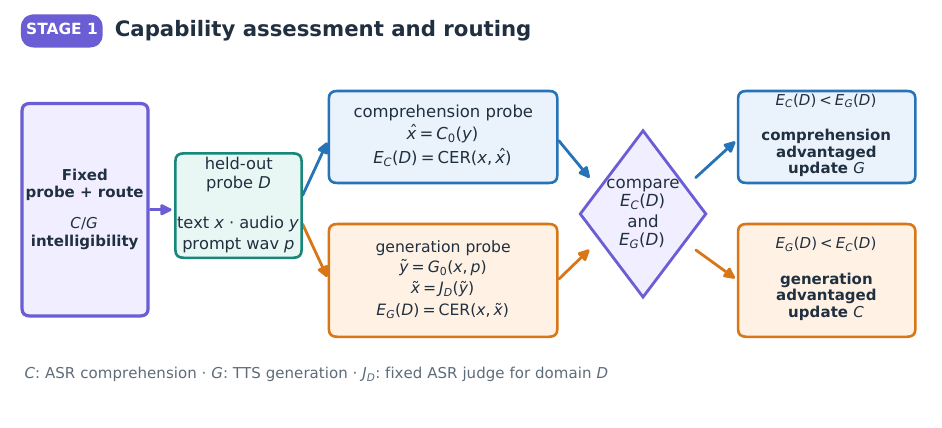}
  \caption{Capability-advantage assessment and routing. ASR and TTS are
  compared through text intelligibility, and the counterpart of the
  lower-error module is selected for update.}
  \label{fig:capability}
\end{figure}

The routing component records $E_C(D)$ and $E_G(D)$ in the advantage table and
applies the following deterministic rule:
\begin{align}
\mathcal D_{C\rightarrow G}
&=\{D\in\mathcal D\mid E_C(D)<E_G(D)\},\\
\mathcal D_{G\rightarrow C}
&=\{D\in\mathcal D\mid E_G(D)<E_C(D)\},
\label{eq:routing}
\end{align}
where $\mathcal D_{C\rightarrow G}$ contains domains with a comprehension
advantage and therefore updates $G$, while $\mathcal D_{G\rightarrow C}$
contains domains with a generation advantage and therefore updates $C$. A tie
activates neither one-way route. Figure~\ref{fig:capability} summarizes the
speech instantiation of this advantage-prior acquisition.

\FloatBarrier

\subsection{Iterative Online Optimization}

After route activation, Stage~2 executes an online loop that resolves the
teacher and student roles, constructs supervision with the corresponding
modules, applies route-specific selection predicates, and updates the selected
student. The following definitions instantiate these operations for
autoregressive ASR and TTS. Let
$x=(x_1,\ldots,x_{|x|})$ be a text-token sequence, $y$ a waveform, $p$ a
prompt waveform, and $z=q(y)$ the discrete speech-token sequence produced by
speech tokenizer $q$. Their factorizations are
\begin{align}
P_{\theta_C}(x\mid y)
&=\prod_{n=1}^{|x|}P_{\theta_C}(x_n\mid x_{<n},y),\\
P_{\theta_G}(z\mid x,p)
&=\prod_{m=1}^{|z|}P_{\theta_G}(z_m\mid z_{<m},x,p),
\label{eq:ar-factorization}
\end{align}
where $\theta_C$ and $\theta_G$ are the trainable parameter subsets of the
comprehension and generation modules, and $x_{<n}$ and $z_{<m}$ are their
autoregressive prefixes.

\subsubsection{Comprehension-supervised mode.}
\begin{figure}[!t]
  \centering
  \includegraphics[width=\linewidth]{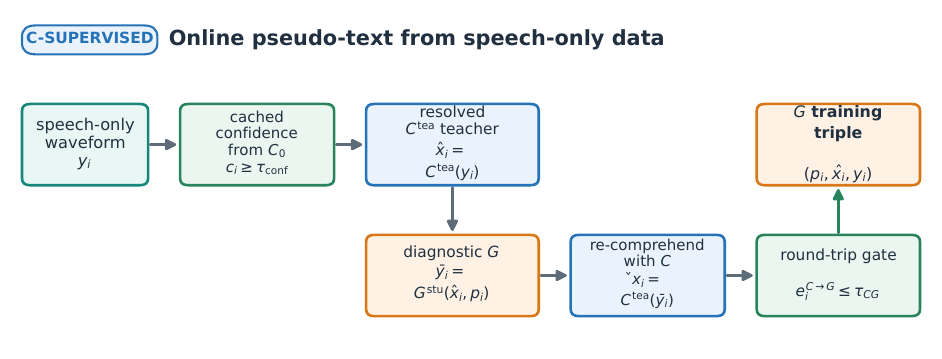}
  \caption{Comprehension-supervised mode. ASR produces pseudo-text, diagnostic
  synthesis and round-trip recognition test its consistency, and the accepted
  original waveforms train TTS.}
  \label{fig:comprehension-supervised}
\end{figure}

For a domain in $\mathcal D_{C\rightarrow G}$, the route resolves $C$ as the
ASR teacher and $G$ as the TTS student. For a real waveform $y_i$, the initial
module $C_0$, with parameters $\theta_C^0$, produces cached pseudo-text
$\hat x_i^0=(\hat x_{i,1}^0,\ldots,\hat x_{i,L_i^0}^0)$, where $L_i^0$ is
the number of valid non-special text tokens.
Figure~\ref{fig:comprehension-supervised} summarizes the route: diagnostic
synthesis verifies pseudo-text consistency, while SFT retains the original
waveform as its acoustic target.

The posterior of token $n$ is
\begin{equation}
\pi_{i,n}=P_{\theta_C^0}(\hat x_{i,n}^0\mid\hat x_{i,<n}^0,y_i),
\end{equation}
and the cached sequence confidence is
\begin{equation}
c_i=\frac{1}{L_i^0}\sum_{n=1}^{L_i^0}\ind[\pi_{i,n}>\rho].
\label{eq:confidence}
\end{equation}
Here $\pi_{i,n}$ is the posterior of cached token $\hat x_{i,n}^0$, $\rho$ is
the token-confidence threshold, $c_i$ is the high-confidence-token fraction,
and $\ind[\cdot]$ is the indicator function. A row passes when
$c_i\ge\tau_{\mathrm{conf}}$, where $\tau_{\mathrm{conf}}$ is the sequence
acceptance threshold.

For a row admitted by the cached-confidence rule, the resolved comprehension
teacher produces online pseudo-text
\begin{equation}
\hat x_i=C^{\mathrm{tea}}(y_i),
\label{eq:pseudotext}
\end{equation}
where $C^{\mathrm{tea}}$ is the resolved ASR teacher and $\hat x_i$ may change
after the teacher is refreshed. The current generation student then produces
a diagnostic waveform that the teacher comprehends again:
\begin{equation}
\bar y_i=G^{\mathrm{stu}}(\hat x_i,p_i),\qquad
\check x_i=C^{\mathrm{tea}}(\bar y_i).
\end{equation}
Here $G^{\mathrm{stu}}$ is the current TTS student, $p_i$ is its VAD-derived or
speaker-pool prompt waveform, $\bar y_i$ is diagnostic speech, and
$\check x_i$ is the round-trip transcript. The round-trip error is
\begin{equation}
e_i^{C\rightarrow G}=
\frac{\Edit(\hat x_i,\check x_i)}{|\hat x_i|},
\label{eq:roundtrip}
\end{equation}
where $\Edit(a,b)$ denotes character-level Levenshtein edit distance between
normalized strings $a$ and $b$. The row passes when
$e_i^{C\rightarrow G}\le\tau_{CG}$, where $\tau_{CG}$ is the round-trip
error threshold.

A generation-training row is accepted only when both predicates hold:
\begin{equation}
\mathcal S_G=\{(p_i,\hat x_i,y_i)\mid
c_i\ge\tau_{\mathrm{conf}},\ e_i^{C\rightarrow G}\le\tau_{CG}\}.
\end{equation}
The diagnostic waveform $\bar y_i$ is used only for selection. The accepted set
$\mathcal S_G$ is then dispatched to the generation-module optimizer. With
original speech tokens $z_i=q(y_i)$, the generation module learns from the
real waveform:
\begin{equation}
\mathcal L_G^{\mathrm{SFT}}=
-\sum_{(p_i,\hat x_i,y_i)\in\mathcal S_G}
\sum_{m=1}^{|z_i|}
\log P_{\theta_G}(z_{i,m}\mid z_{i,<m},\hat x_i,p_i).
\label{eq:generation-sft}
\end{equation}
Here $\mathcal L_G^{\mathrm{SFT}}$ is the generation-side supervised
fine-tuning loss.

\FloatBarrier

\subsubsection{Generation-supervised mode.}
\begin{figure}[!h]
  \centering
  \includegraphics[width=\linewidth]{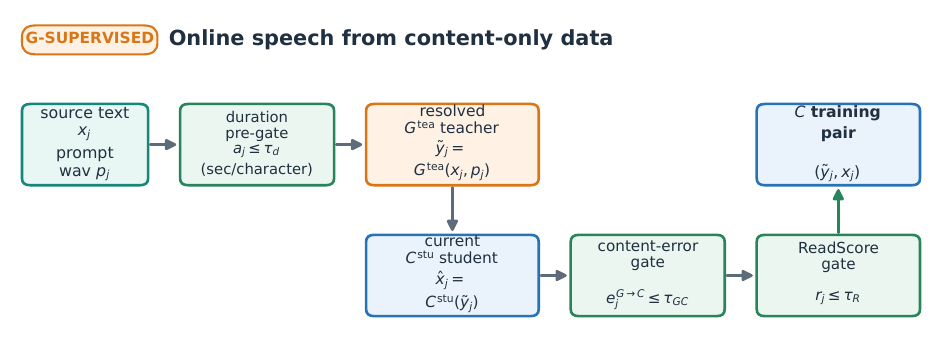}
  \caption{Generation-supervised mode. TTS renders content-only samples into
  speech, and duration, student-intelligibility, and ReadScore rules select
  synthetic speech--content pairs for ASR training.}
  \label{fig:generation-supervised}
\end{figure}

For a domain in $\mathcal D_{G\rightarrow C}$, the route resolves $G$ as the
TTS teacher and $C$ as the ASR student. A duration gate is applied to each
source text $x_j$ and prompt waveform $p_j$ before online synthesis.
Figure~\ref{fig:generation-supervised} summarizes how the
route renders content-only rows into speech and selects synthetic
speech--content pairs for comprehension training.

Define
\begin{equation}
n_j^{\mathrm{char}}=|x_j|,\qquad
a_j=\frac{\dur(\tilde y_j^0)}{\max(n_j^{\mathrm{char}},1)},
\label{eq:length-ratio}
\end{equation}
where $n_j^{\mathrm{char}}$ is the normalized source-text character count,
$\tilde y_j^0=G_0(x_j,p_j)$ is initial diagnostic synthesis, $\dur(\cdot)$
returns seconds, and $a_j$ is seconds per character. The prompt $p_j$ is the
VAD-derived or speaker-pool acoustic context associated with row $j$. A row
passes when $a_j\le\tau_d$, where $\tau_d$ is the duration-ratio threshold.

For each row admitted by the duration rule, the resolved generation teacher
produces online speech, which the current comprehension student decodes:
\begin{equation}
\tilde y_j=G^{\mathrm{tea}}(x_j,p_j),\qquad
\hat x_j=C^{\mathrm{stu}}(\tilde y_j).
\label{eq:generation-comprehension-chain}
\end{equation}
Here $G^{\mathrm{tea}}$ is the resolved TTS teacher, $\tilde y_j$ is its
synthetic waveform, $C^{\mathrm{stu}}$ is the current ASR student, and
$\hat x_j$ is the student hypothesis. Their content error is
\begin{equation}
e_j^{G\rightarrow C}=
\frac{\Edit(x_j,\hat x_j)}{|x_j|}.
\label{eq:student-cer}
\end{equation}
Here again, $\Edit(a,b)$ is character-level Levenshtein edit distance between
normalized strings $a$ and $b$. The row passes when
$e_j^{G\rightarrow C}\le\tau_{GC}$, where $\tau_{GC}$ is the student-error
threshold.

We apply the published ReadScore rule \cite{li2026read} and denote its output by
\begin{equation}
r_j=\ReadScore(\tilde y_j,x_j,\hat x_j;G^{\mathrm{tea}}).
\label{eq:readscore}
\end{equation}
Here $r_j$ is the reference-free discrepancy returned for waveform
$\tilde y_j$, intended text $x_j$, and hypothesis $\hat x_j$, using the
resolved autoregressive generation model as scorer. The row passes when
$r_j\le\tau_R$, where $\tau_R$ is the ReadScore acceptance threshold.

A synthetic waveform is accepted only when all three predicates hold. The
resulting set defines the following comprehension-training objective:
\begin{equation}
\mathcal S_C=\{(\tilde y_j,x_j)\mid a_j\le\tau_d,\
e_j^{G\rightarrow C}\le\tau_{GC},\ r_j\le\tau_R\},
\end{equation}
\begin{equation}
\mathcal L_C^{\mathrm{SFT}}=
-\sum_{(\tilde y_j,x_j)\in\mathcal S_C}\sum_{n=1}^{|x_j|}
\log P_{\theta_C}(x_{j,n}\mid x_{j,<n},\tilde y_j),
\label{eq:comprehension-sft}
\end{equation}
where $\mathcal L_C^{\mathrm{SFT}}$ is the comprehension-side supervised
fine-tuning loss.

\FloatBarrier

\begin{figure}[!t]
  \centering
  \includegraphics[width=\linewidth]{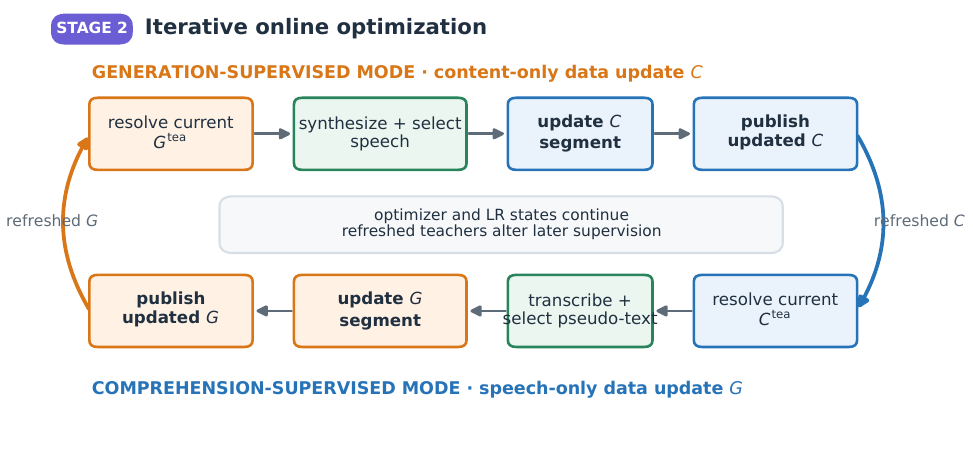}
  \caption{Iterative online optimization. Current teacher checkpoints drive
  route-specific supervision construction and selection, while optimizer and
  learning-rate states continue across updates.}
  \label{fig:cycles}
\end{figure}

After each selected update, optimizer and learning-rate states continue, the
latest relevant checkpoints are resolved, and the stopping condition is
checked. Supervision construction and model-dependent selection use the newly
resolved teacher, so later cycles can induce a different supervision
distribution from earlier cycles.
Figure~\ref{fig:cycles} shows the resulting complete loop: the upper route
updates $C$, and the reversed lower route uses the refreshed $C$ to update $G$
without crossing exchange arrows.

\FloatBarrier

\subsection{Deviation-Constrained Optimization via On-Policy Distillation}

We apply OPD \cite{agarwal2024opd} as a preservation constraint when either
autoregressive module is selected for update. Let
$m\in\{C,G\}$ identify the module being updated, $u_m^o$ a conditioning input
from its preservation domain, and
\begin{equation}
v_m^S=(v_{m,1}^S,\ldots,v_{m,L_m}^S)
\sim P_{\theta_m}(\cdot\mid u_m^o)
\end{equation}
the output trajectory sampled from the current module. On every
student-generated prefix $v_{m,<\ell}^S$, the frozen original module with
parameters $\theta_m^0$ and the current module produce next-token distributions.
We minimize
\begin{equation}
\mathcal L_{\mathrm{OPD}}^{(m)}=
\frac{1}{L_m}\sum_{\ell=1}^{L_m}
D_{\mathrm{KL}}\!\left(
P_{\theta_m^0}(\cdot\mid u_m^o,v_{m,<\ell}^S)\,\|\,
P_{\theta_m}(\cdot\mid u_m^o,v_{m,<\ell}^S)
\right),
\label{eq:opd}
\end{equation}
\begin{equation}
\mathcal L_m=\mathcal L_m^{\mathrm{SFT}}
+\lambda_m\mathcal L_{\mathrm{OPD}}^{(m)}.
\end{equation}
Here $L_m$ is the sampled trajectory length, $\theta_m^0$ denotes the original
teacher parameters, $D_{\mathrm{KL}}$ is teacher-to-student KL divergence,
$\mathcal L_m^{\mathrm{SFT}}$ is the selected module's main supervised loss,
and $\lambda_m$ is the deviation-constraint weight. For $m=C$, the trajectory
contains content tokens conditioned on an acoustic observation; for $m=G$, it
contains speech tokens conditioned on content and the applicable acoustic
context. In the reported experiment, this formulation is used for both
modules. Each updated module returns to the online loop, and the final optimized
pair is returned after the configured stopping budgets are reached.

\section{Experiments}

\subsection{Experimental Setup}
\label{sec:setup}

\subsubsection{Models and one-sided data.}
The initial modules are Whisper large-v3 \cite{radford2023whisper} and the
official CosyVoice2 checkpoint \cite{du2024cosyvoice2}.
Table~\ref{tab:data} summarizes the data used for training, capability probing,
and final evaluation. Training uses RAMC text only \cite{yang2022ramc} and
CV-JA and JVS speech only \cite{ardila2020commonvoice,takamichi2019jvs}.
For Japanese Common Voice, training uses CV-JA validated after excluding all
utterances assigned to the official dev and test splits, whereas testing uses
CV-JA test GE5. GE5 denotes utterances longer than 5~s.
In JVS, \texttt{jvs001}--\texttt{jvs100} denote speaker IDs; speakers
\texttt{jvs001}--\texttt{jvs090} form the training split, while the
speaker-disjoint \texttt{jvs091}--\texttt{jvs100} split is reserved for
Stage-1 probing.
The RAMC Stage-1 probe uses the official dev sessions, which are disjoint from
the official test sessions used for final evaluation.

\begin{table}[!ht]
\centering
\caption{Data inventory and TTS prompt sources.}
\label{tab:data}
\scriptsize
\setlength{\tabcolsep}{2.1pt}
\begin{tabularx}{\linewidth}{@{}l
>{\raggedright\arraybackslash}p{0.20\linewidth}
>{\raggedright\arraybackslash}p{0.075\linewidth}rrr
>{\raggedright\arraybackslash}X@{}}
\toprule
Split & Dataset & Side & Rows & Orig. h & Syn. h & TTS prompt source \\
\midrule
\multirow{3}{*}{Train}
 & RAMC (ZH) & text & 178,904 & 133.673 & 185.255
 & real-speaker pool, 3--7 s \\
 & CV-JA validated & audio & 94,303 & 188.095 & --
 & same utterance, random VAD 2--3 s \\
 & JVS jvs001--090 & audio & 12,954 & 26.317 & --
 & same utterance, random VAD 2--3 s \\
\midrule
\multirow{2}{*}{Dev}
 & JVS jvs091--100 & paired & 1,447 & 2.972 & --
 & same utterance, first VAD 2--3 s \\
 & RAMC dev GE5 & paired & 1,689 & 2.953 & --
 & same utterance, first VAD 2--3 s \\
\midrule
\multirow{2}{*}{Test}
 & CV-JA test GE5 & paired & 3,163 & 6.571 & --
 & same utterance, first VAD 2--3 s \\
 & RAMC test GE5 & paired & 3,634 & 6.433 & --
 & same utterance, first VAD 2--3 s \\
\bottomrule
\end{tabularx}
\end{table}

\subsubsection{Selection, optimization, and stopping.}
Each cycle starts from 512 source rows. For updating $G$, the token and
sequence confidence thresholds are $\rho=0.8$ and
$\tau_{\mathrm{conf}}=0.8$; the round-trip threshold is $\tau_{CG}=0.1$.
For updating $C$, the duration threshold is $\tau_d=0.8$ s/character and the
student-error threshold is $\tau_{GC}=0.2$. We use ReadScore with its internal
discrepancy parameter set to 0.1 and outer acceptance threshold
$\tau_R=0.4$.

Whisper is trained in BF16 with 8-way FSDP and effective batch 64. Its peak
learning rate is $8\times10^{-6}$, with 20 warm-up steps and a multiplicative
post-warm-up decay of 0.995 for a total of 200 optimizer steps. CosyVoice2 SFT
updates only its autoregressive LLM; the flow and vocoder remain frozen. The
LLM uses FP16, 8-way DeepSpeed Stage~2, dynamic batching capped at 480 frames,
a peak learning rate of $2\times10^{-5}$, 60 warm-up steps, and decay 0.998
for 600 optimizer steps. In the online runs, optimizer and scheduler states
continue across cycles without restarting warm-up. For each module, the online
run and its matched one-shot static reference start from the same official
initialization and use the same optimizer-step budget and learning-rate
schedule. The static references globally shuffle the preconstructed Chinese
and Japanese supervision once, rather than regenerating and selecting it
across cycles.

Bilateral OPD is applied. When updating $C$, we use the frozen
official Whisper as teacher, 64 Japanese audio samples and four insertion
events per cycle, $\lambda_C=0.5$, greedy on-policy generation, and at most 256
new text tokens. When updating $G$, we use the frozen official CosyVoice2 as teacher,
64 Mandarin texts and eight insertion events per cycle, with
$\lambda_G=0.8$. For counterpart exchange, the first cycle on either side uses
the official model. Thereafter, at starting steps $s_G$ and $s_C$, we use a
Whisper checkpoint covering $\lfloor s_G/3\rfloor$ for the $G$ route and a
CosyVoice2 checkpoint covering $\min(3s_C,600)$ for the $C$
route. Final TTS evaluation uses Qwen3-ASR-1.7B CER, speaker cosine, and UTMOS; the
Stage-1 routing criterion uses intelligibility CER alone.

\subsection{Advantage Prior and Activated Modes}

Table~\ref{tab:capability} reports one Stage-1 probe for each activated mode.
On the Japanese JVS probe, $E_C=5.94\%$ is obtained by Whisper large-v3 on the
real waveforms, whereas the generation-side judge $J_D$ is
Qwen3-ASR-1.7B and yields $E_G=12.09\%$ on CosyVoice2 speech. The resulting
comprehension advantage selects the generation module for update. On the
disjoint official RAMC dev probe, generation has lower CER and the routing
rule selects the comprehension module for update.

\begin{table}[h]
\centering
\caption{Intelligibility-based advantage prior (CER, \%; lower is better).}
\label{tab:capability}
\small
\setlength{\tabcolsep}{3.5pt}
\begin{tabular}{>{\raggedright\arraybackslash}p{0.27\linewidth}rr
>{\raggedright\arraybackslash}p{0.20\linewidth}
>{\raggedright\arraybackslash}p{0.18\linewidth}}
\toprule
Stage-1 probe & $E_C$ & $E_G$ & Advantaged module & Updated module \\
\midrule
Japanese (JVS) & 5.94 & 12.09 & $C$ (ASR) & $G$ (TTS) \\
Chinese (RAMC dev) & 14.27 & 9.46 & $G$ (TTS) & $C$ (ASR) \\
\bottomrule
\end{tabular}
\end{table}

\subsection{Quality of the Optimized Modules}

The online run terminates at Whisper step 200 after 37 data cycles and at
CosyVoice2 step 600 after 14 data cycles. Table~\ref{tab:asr-quality} reports
comprehension quality on the final tests. Relative to the official checkpoint,
the optimized Whisper improves CV-JA from 15.23\% to 15.10\% CER and RAMC from
14.18\% to 10.77\%. Under the matched step and learning-rate schedule, online
optimization also improves over the one-shot static construction on both CV-JA
(18.75\% to 15.10\%) and RAMC (11.63\% to 10.77\%). This is a system-level
comparison: teacher refresh, online selection, and preservation data jointly
change the supervision, so it supports the complete online configuration
rather than an isolated cycling-only effect. In Tables~\ref{tab:asr-quality}
and~\ref{tab:tts-quality}, ``(OPD)'' marks the preservation source used only by
the OPD lane.

\begin{table}[t]
\centering
\caption{Whisper large-v3 comprehension quality. CER is in percent; lower is
better. Bold compares iterative online optimization with the matched-step
static reference.}
\label{tab:asr-quality}
\scriptsize
\setlength{\tabcolsep}{2.5pt}
\begin{tabularx}{\linewidth}{@{}>{\raggedright\arraybackslash}X>{\raggedright\arraybackslash}Xrrr@{}}
\toprule
Optimization scheme & Training data & Steps & CV-JA CER & RAMC CER \\
\midrule
Official checkpoint & -- & -- & 15.23 & 14.18 \\
One-shot static merge & ZH text + JA audio & 200 & 18.75 & 11.63 \\
Iterative online & ZH text + JA audio (OPD) & 200 & \textbf{15.10} & \textbf{10.77} \\
\bottomrule
\end{tabularx}
\end{table}

Table~\ref{tab:tts-quality} reports final-test generation quality. Relative to
the official checkpoint, online CosyVoice2 improves CV-JA CER from 14.85\% to
11.18\% and RAMC CER from 4.78\% to 4.04\%. Against the matched-step static
reference, however, iterative online optimization is close but slightly worse on CV-JA
(11.18\% versus 11.03\%) and trails on RAMC (4.04\% versus 1.93\%). Speaker
cosine and UTMOS remain close to the static construction, indicating that the
RAMC difference is primarily reflected by intelligibility CER rather than a
broad collapse of the generation model.

\begin{table}[t]
\centering
\caption{CosyVoice2 generation quality. CER is computed by Qwen3-ASR-1.7B and
is in percent; Spk is prompt-to-output speaker cosine. Static and iterative
online runs use the same 600-step learning-rate schedule.}
\label{tab:tts-quality}
\scriptsize
\setlength{\tabcolsep}{1.8pt}
\begin{tabularx}{\linewidth}{@{}>{\raggedright\arraybackslash}X>{\raggedright\arraybackslash}Xrrrrrr@{}}
\toprule
\multicolumn{2}{c}{} & \multicolumn{3}{c}{CV-JA} & \multicolumn{3}{c}{RAMC} \\
\cmidrule(lr){3-5}\cmidrule(l){6-8}
Optimization scheme & Training data
  & CER$\downarrow$ & Spk$\uparrow$ & UTMOS$\uparrow$
  & CER$\downarrow$ & Spk$\uparrow$ & UTMOS$\uparrow$ \\
\midrule
Official checkpoint & --
  & 14.85 & .635 & 2.741
  & 4.78 & .670 & 2.972 \\
One-shot static merge & ZH text + JA audio
  & 11.03 & .700 & 3.305
  & 1.93 & .663 & 3.048 \\
Iterative online & JA audio + ZH text (OPD)
  & 11.18 & .702 & 3.290
  & 4.04 & .673 & 2.984 \\
\bottomrule
\end{tabularx}
\end{table}

The online-versus-static comparison is therefore module dependent. The
$G$-supervised route refreshes complete synthetic waveforms used to train
Whisper, whereas the $C$-supervised route retains real acoustic targets and
refreshes only pseudo-text and row selection. Moreover, the resolved Whisper
teacher changes only slightly on CV-JA, from 15.23\% to 15.10\% CER, limiting
the potential gain from later pseudo-text refreshes. Finally, online
CosyVoice2 receives Japanese SFT with a smaller Mandarin OPD preservation
lane, whereas the static merge directly includes the full synthetic RAMC
supervision. These are observational explanations rather than isolated causal
findings because the static and online systems also differ in data
construction, selection, and OPD.

\section{Conclusion}

We presented a two-stage capability-guided iterative online optimization
pipeline for the peripheral comprehension and generation modules of a
cascaded spoken dialogue system. A domain-wise intelligibility probe first
obtains a capability prior and selects either comprehension-supervised
optimization from speech-only data or generation-supervised optimization from
content-only data. In the latter route, the generation module renders content
into speech, and the accepted synthetic speech--content pairs train
comprehension. Across cycles, the two routes construct and select supervision
online, exchange resolved checkpoints, and apply bilateral OPD.

Both Whisper large-v3 and CosyVoice2 reduce CER relative to their official
initializations on both reported tests. Under matched optimizer-step budgets
and learning-rate schedules, iterative online Whisper also outperforms its
one-shot static reference on both tests. Iterative online CosyVoice2 is close
to but slightly worse than the static reference on Japanese and trails it on
Mandarin. Thus, the benefit of checkpoint-refreshed online supervision is
clear for comprehension but remains module dependent; improving the generation
route is an important direction for future work.


\begin{thebibliography}{15}

\bibitem{agarwal2024opd}
Agarwal, R., Vieillard, N., Zhou, Y., Stanczyk, P., Ramos Garea, S.,
Geist, M., Bachem, O.: On-policy distillation of language models:
Learning from self-generated mistakes. In: International Conference
on Learning Representations (ICLR) (2024).
\url{https://proceedings.iclr.cc/paper_files/paper/2024/hash/5be69a584901a26c521c2b51e40a4c20-Abstract-Conference.html}

\bibitem{ardila2020commonvoice}
Ardila, R., Branson, M., Davis, K., Kohler, M., Meyer, J., Henretty, M.,
Morais, R., Saunders, L., Tyers, F.M., Weber, G.: Common Voice: A
massively-multilingual speech corpus. In: Proceedings of the Twelfth
Language Resources and Evaluation Conference (LREC 2020),
pp. 4218--4222. European Language Resources Association, Marseille
(2020). \url{https://aclanthology.org/2020.lrec-1.520/}

\bibitem{baskar2019semisupervised}
Baskar, M.K., Watanabe, S., Astudillo, R., Hori, T., Burget, L.,
\v{C}ernock\'y, J.: Semi-supervised sequence-to-sequence ASR using
unpaired speech and text. In: Interspeech 2019, pp. 3790--3794.
ISCA (2019). \doi{10.21437/Interspeech.2019-3167}

\bibitem{du2024cosyvoice2}
Du, Z., Wang, Y., Chen, Q., Shi, X., Lv, X., Zhao, T., Gao, Z., Yang, Y.,
Gao, C., Wang, H., Yu, F., Liu, H., Sheng, Z., Gu, Y., Deng, C., Wang, W.,
Zhang, S., Yan, Z., Zhou, J.: CosyVoice 2: Scalable streaming speech
synthesis with large language models. arXiv preprint arXiv:2412.10117
(2024). \doi{10.48550/arXiv.2412.10117}

\bibitem{huang2024audiogpt}
Huang, R., Li, M., Yang, D., Shi, J., Chang, X., Ye, Z., Wu, Y., Hong, Z.,
Huang, J., Liu, J., Ren, Y., Zou, Y., Zhao, Z., Watanabe, S.: AudioGPT:
Understanding and generating speech, music, sound, and talking head.
Proceedings of the AAAI Conference on Artificial Intelligence
\textbf{38}(21), 23802--23804 (2024).
\doi{10.1609/aaai.v38i21.30570}

\bibitem{kuan2024speechcopilot}
Kuan, C.-Y., Yang, C.-K., Huang, W.-P., Lu, K.-H., Lee, H.-Y.:
Speech-Copilot: Leveraging large language models for speech processing
via task decomposition, modularization, and program generation. In:
2024 IEEE Spoken Language Technology Workshop (SLT), pp. 1060--1067.
IEEE (2024). \doi{10.1109/SLT61566.2024.10832184}

\bibitem{li2026read}
Li, Z., Wang, H., Guo, Y., Li, B., Chen, X., Yu, K.: Read what you hear:
Reference-free hypotheses evaluation with acoustic discrepancy. arXiv
preprint arXiv:2606.04680 (2026).
\doi{10.48550/arXiv.2606.04680}

\bibitem{liu2025xtalk}
Liu, Z., Duan, Y., Wang, M., Feng, P., Zhang, H., Xing, X., Shan, Y.,
Zhu, H., Dai, Y., Lu, C., Qiu, X., Xie, L., Wang, L., Yan, N., Zheng, Z.,
Ma, Z., Yu, K., Chen, X.: X-Talk: On the underestimated potential of
modular speech-to-speech dialogue system. arXiv preprint
arXiv:2512.18706 (2025). \doi{10.48550/arXiv.2512.18706}

\bibitem{maben2025aura}
Maben, L.M., Lakshmy, G.G., Radhakrishnan, S., Arora, S., Watanabe, S.:
AURA: Agent for understanding, reasoning, and automated tool use in
voice-driven tasks. In: 2025 IEEE Automatic Speech Recognition and
Understanding Workshop (ASRU), pp. 1--4. IEEE (2025).
\doi{10.1109/ASRU65441.2025.11434725}

\bibitem{qi2023speechain}
Qi, H., Novitasari, S., Tjandra, A., Sakti, S., Nakamura, S.: SpeeChain:
A speech toolkit for large-scale machine speech chain. arXiv preprint
arXiv:2301.02966 (2023). \doi{10.48550/arXiv.2301.02966}

\bibitem{radford2023whisper}
Radford, A., Kim, J.W., Xu, T., Brockman, G., McLeavey, C., Sutskever, I.:
Robust speech recognition via large-scale weak supervision. In: Krause,
A., Brunskill, E., Cho, K., Engelhardt, B., Sabato, S., Scarlett, J. (eds.)
Proceedings of the 40th International Conference on Machine Learning,
Proceedings of Machine Learning Research, vol. 202, pp. 28492--28518.
PMLR (2023). \url{https://proceedings.mlr.press/v202/radford23a.html}

\bibitem{ren2019almost}
Ren, Y., Tan, X., Qin, T., Zhao, S., Zhao, Z., Liu, T.-Y.: Almost
unsupervised text to speech and automatic speech recognition. In:
Chaudhuri, K., Salakhutdinov, R. (eds.) Proceedings of the 36th
International Conference on Machine Learning, Proceedings of Machine
Learning Research, vol. 97, pp. 5410--5419. PMLR (2019).
\url{https://proceedings.mlr.press/v97/ren19a.html}

\bibitem{takamichi2019jvs}
Takamichi, S., Mitsui, K., Saito, Y., Koriyama, T., Tanji, N.,
Saruwatari, H.: JVS corpus: Free Japanese multi-speaker voice corpus.
arXiv preprint arXiv:1908.06248 (2019).
\doi{10.48550/arXiv.1908.06248}

\bibitem{tjandra2017speechchain}
Tjandra, A., Sakti, S., Nakamura, S.: Listening while speaking: Speech
chain by deep learning. In: 2017 IEEE Automatic Speech Recognition and
Understanding Workshop (ASRU), pp. 301--308. IEEE (2017).
\doi{10.1109/ASRU.2017.8268950}

\bibitem{yang2022ramc}
Yang, Z., Chen, Y., Luo, L., Yang, R., Ye, L., Cheng, G., Xu, J., Jin, Y.,
Zhang, Q., Zhang, P., Xie, L., Yan, Y.: Open source MagicData-RAMC: A
rich annotated Mandarin conversational (RAMC) speech dataset. In:
Interspeech 2022, pp. 1736--1740. ISCA (2022).
\doi{10.21437/Interspeech.2022-729}

\end{thebibliography}
\end{document}